\documentclass[a4paper,aps,prd,10pt,preprintnumbers,showpacs,twocolumn,superscriptaddress,nofootinbib,amsmath,amssymb]{revtex4-1}
\usepackage{graphicx}
\usepackage{cmap}
\usepackage[utf8]{inputenc}
\usepackage[T1]{fontenc}

\begin{document}
\title{Scattering and Absorption of Standard Model Fields by  Brane-Localized Schwarzschild--de Sitter Black Holes}
\author{Alexey Dubinsky}
\email{\texttt{dubinsky@ukr.net}}
\affiliation{University of Seville, Seville, Spain}

\begin{abstract}
We investigate the propagation and absorption of Standard Model fields---scalar, electromagnetic, and Dirac---on a 3+1-dimensional brane embedded in a higher-dimensional Schwarzschild--de Sitter (SdS) spacetime. Using the effective four-dimensional projection of the Tangherlini metric, we compute grey-body factors (GBFs) and absorption cross-sections for each spin sector by means of the sixth-order WKB method and, independently, via the recently proposed correspondence between quasinormal modes (QNMs) and transmission coefficients. The results demonstrate that the cosmological constant and field mass crucially affect the transmission probabilities: increasing $\Lambda$ lowers the potential barrier and enhances the transparency of the geometry, while a larger field mass $\mu$ suppresses low-frequency emission and shifts the absorption spectrum to higher energies. For all fields, the QNM--GBF correspondence proves reliable to within about one percent for multipoles $\ell \ge 2$, while the correspondence remains less accurate for the lowest multipoles. The total absorption cross-sections exhibit the expected transition from the low-frequency suppression to the geometric-optics regime. Overall, these findings provide quantitative insight into the interplay between dimensionality, cosmological expansion, and black-hole radiation on the brane.\end{abstract}

\maketitle

\section{Introduction}

Black holes in higher-dimensional spacetimes have long been a central subject of theoretical research, motivated by developments in string theory, braneworld cosmology, and the gauge/gravity correspondence. In models with large or warped extra dimensions, our observable Universe is often represented as a $(3+1)$-dimensional brane embedded in a higher-dimensional bulk spacetime~\cite{Randall:1999ee,Arkani-Hamed:1998jmv}. In this context, gravity propagates in the full higher-dimensional manifold, while Standard Model fields are confined to the brane. As a consequence, black holes formed by localized matter sources appear to observers on the brane as effectively four-dimensional objects, although their geometry retains imprints of the higher-dimensional structure. The projection of higher-dimensional black holes onto the brane thus provides a powerful setting to study how extra-dimensional effects can modify classical black-hole observables, including Hawking radiation spectra, quasinormal oscillations, and grey-body factors (GBFs)~\cite{Kanti:2004nr,Maartens:2003tw,Seahra:2004fg,Zhidenko:2008fp,Bronnikov:2019sbx,Ishihara:2008re,Cardoso:2005vb,Kanti:2002nr,Fitzpatrick:2006cd,Kanti:2005xa,Zinhailo:2024jzt,Kanti:2006ua,Kanti:2009sn,Stuchlik:2025ezz}.

When a black hole emits Hawking radiation, the spectrum is not purely thermal because the outgoing quanta are partially scattered by the curvature-induced potential surrounding the horizon. The resulting transmission probability is encoded in the grey-body factors ~\cite{Page:1976df,Page:1976ki,Kanti:2002nr,Cardoso:2005mh}, which effectively act as frequency-dependent filters on the emitted radiation (see~\cite{Antoniou:2025bvg,Konoplya:2025ixm,Konoplya:2025uta} for recent examples). The GBFs thus provide an essential link between the microscopic emission process at the horizon and the observable flux measured at infinity. They also determine the total absorption cross-section, which characterizes how the black hole interacts with external radiation fields. For asymptotically de Sitter (dS) spacetimes, this analysis becomes more subtle due to the presence of a cosmological horizon, which modifies the boundary conditions and the scattering structure~\cite{Brady:1996za,Molina:2003ff,Choudhury:2003wd}. In such backgrounds, the transmission and reflection probabilities are defined between the black-hole and cosmological horizons, and the grey-body factors depend sensitively on the effective potential shape between them.

Different spin fields probe the geometry in distinct ways. Scalar ($s=0$), electromagnetic ($s=1$), and Dirac ($s=1/2$) fields are particularly useful test cases because their perturbation equations can be cast in a Schrödinger-like form with well-defined effective potentials~\cite{Cho:2003qe,Konoplya:2011qq,Kanti:2004nr}. These potentials determine the scattering properties, and hence the grey-body factors, through the corresponding wave equations in the tortoise coordinate. Comparison between the various spin cases reveals how the spin–curvature coupling affects the transmission of radiation across the gravitational barrier, especially in the presence of a positive cosmological constant. It is worth recalling that an effective cosmological constant may naturally arise in a broad class of $f(R)$ theories even without an explicit $\Lambda$ term in the Lagrangian, as constant-curvature solutions of the form $R = R_0 = \mathrm{const}$ play the role of (anti--)de Sitter vacua~\cite{Sotiriou:2008rp,DeFelice:2010aj,Nojiri:2017ncd,Starobinsky:2007hu}.

A further motivation for studying grey-body factors stems from their recently established  correspondence with quasinormal modes (QNMs). In the eikonal regime, the transmission coefficient can be directly related to the complex QNM frequencies via the WKB expansion, leading to an approximate but often remarkably accurate connection between the spectral and scattering properties of black holes~\cite{Konoplya:2024lir}. However, this correspondence is not universal: it holds most accurately for single-barrier potentials of asymptotically flat black holes but can fail or become merely approximate when the WKB approximation does not work properly~\cite{Lutfuoglu:2025ljm,Lutfuoglu:2025ohb,Dubinsky:2025nxv,Bolokhov:2024otn,Lutfuoglu:2025blw}. Testing the robustness of this relation in various geometries and for fields of different spin remains an important question \cite{Lutfuoglu:2025hjy,Han:2025cal,Dubinsky:2025nxv,Malik:2024cgb,Lutfuoglu:2025blw,Lutfuoglu:2025ldc,Bolokhov:2024otn,Malik:2025erb,Dubinsky:2024vbn}.

Grey-body factors and Hawking radiation of scalar fields in higher-dimensional asymptotically de Sitter black holes have been investigated in~\cite{Kanti:2014dxa,Pappas:2016ovo,Kanti:2017ubd,Pappas:2017kam}, while quasinormal modes of brane-localized Schwarzschild--de Sitter black holes were analyzed in~\cite{Kanti:2005xa}. Nevertheless, to the best of our knowledge, a systematic study of grey-body factors and absorption cross-sections for the full set of Standard Model fields has not yet been carried out. Furthermore, within the framework of the gauge/gravity correspondence, grey-body factors play a crucial role in holographic transport phenomena. They characterize how bulk field excitations propagate across the black-hole potential barrier, thereby controlling the frequency-dependent response — such as the optical conductivity — of the dual strongly coupled plasma or holographic superconductor. In particular, the poles of the transmission coefficient are directly related to the electromagnetic response functions on the boundary, providing a bridge between black-hole scattering and linear transport phenomena in strongly coupled systems ~\cite{Horowitz:2008bn,Herzog:2009xv,Konoplya:2009hv}.
  
In this work, we study the grey-body factors and absorption cross-sections for scalar, electromagnetic, and Dirac test fields in the spacetime of a higher dimensional Schwarzschild–de Sitter black hole projected onto a $(3+1)$-dimensional brane. Using both the WKB method and the correspondence between quasinormal modes and grey-body factors \cite{Konoplya:2024lir}, we examine how the cosmological constant and the spin of the field influence the transmission probabilities and absorption spectra. Furthermore, we test the validity of the quasinormal mode–grey-body factor correspondence in this brane-induced de Sitter background, identifying the parameter ranges in which the relation remains reliable. This analysis not only extends previous results for asymptotically flat and higher-dimensional cases but also provides new insight into quantum radiation and wave propagation in brane-world geometries with a positive cosmological constant.

The paper is organized as follows. In Sec.~II we introduce the higher-dimensional Schwarzschild--de Sitter geometry, its projection onto the 3+1-dimensional brane, and the resulting master equations for scalar, electromagnetic, and Dirac perturbations together with their effective potentials. Section~III describes the computation of grey-body factors using the WKB formalism and analyzes their dependence on the spin of the field, the cosmological constant, and the field mass. In Sec.~IV we test the correspondence between grey-body factors and quasinormal modes, quantifying its accuracy across different multipoles and background parameters. Section~V discusses the total absorption cross-sections and their physical interpretation. Finally, Sec.~VI summarizes the main results and outlines possible extensions of the analysis.

\section{Black-hole metric projected on the brane and wave-like equations.}\label{sec:background}

The higher-dimensional generalization of the Schwarzschild solution was first obtained by Tangherlini~\cite{Tangherlini:1963bw}, providing the natural static and spherically symmetric black hole spacetime in \(D\) dimensions.  
In this framework, the geometry of the bulk is described by the line element
\begin{equation}
ds^2 = -f(r)\, dt^2 + \frac{dr^2}{f(r)} + r^2\, d\Omega_{D-2}^2,
\label{eq:fullmetric}
\end{equation}
where \( d\Omega_{D-2}^2 \) represents the metric of a unit \((D-2)\)-sphere.  
The metric function is,  
\begin{equation}
f(r) = 1 - \frac{2M}{r^{D-3}} - \frac{2 \Lambda r^2}{(D-1)(D-2)},
\label{eq:metricfunction}
\end{equation}
with \( M \) denoting the black-hole mass and \( \Lambda \) the cosmological constant.  
For \( D=4 \) and \( \Lambda = 0 \), the classical Schwarzschild spacetime is recovered, while a non-zero \( \Lambda \) yields the higher-dimensional analogue of the Schwarzschild–de~Sitter black hole.

In brane-world scenarios inspired by string theory, the observable universe is modeled as a 3+1-dimensional hypersurface (the ``brane’’) embedded in a higher-dimensional bulk. Standard-Model fields are confined to this brane, while gravity propagates in all \(D\) dimensions.  
When the extra dimensions are large compared to the black-hole radius, the gravitational field near the brane can be approximated by the higher-dimensional Tangherlini metric. To analyze field propagation perceived by a brane observer, one projects the bulk geometry onto a four-dimensional submanifold by fixing the additional angular coordinates.  
This effectively replaces the higher-dimensional angular sector by the ordinary two-sphere:
\begin{equation}
d\Omega_{2}^{2} = d\theta^{2} + \sin^{2}\theta\, d\phi^{2}.
\end{equation}

The induced metric on the brane therefore takes the form
\begin{equation}
ds^{2} = -f(r)\,dt^{2} + \frac{dr^{2}}{f(r)} + r^{2}\bigl(d\theta^{2} + \sin^{2}\theta\, d\phi^{2}\bigr),
\label{eq:projectedmetric}
\end{equation}
where the function \(f(r)\) remains that of the higher-dimensional bulk, as given by Eq.~\eqref{eq:metricfunction}, and thus encodes the dependence on \(D\), \(M\), and \(\Lambda\).  
Although this projection does not constitute a self-consistent four-dimensional solution of Einstein’s equations, it accurately reproduces the effective gravitational background experienced by fields localized on the brane. This simplified approach has been successfully employed in numerous studies of black-hole perturbations and Hawking radiation in brane-world contexts~\cite{Berti:2003yr,Kanti:2004nr,Kanti:2005xa,Kanti:2006ua,Konoplya:2017ymp}.


Throughout this work, we adopt geometrized units in which \( M = 1 \), so that all dimensional quantities are expressed in terms of the black-hole mass scale. Introducing the tortoise coordinate $r_{*}$ via
\begin{equation}
  \frac{dr_{*}}{dr}=\frac{1}{f(r)},
  \label{eq:tortoise}
\end{equation}
all spin-$s$ test fields considered here reduce (after separation of variables) to a one-dimensional Schr\"odinger-type equation
\begin{equation}
  \frac{d^{2}\Psi}{dr_{*}^{2}}+\bigl(\omega^{2}-V_{s}(r)\bigr)\Psi=0,
  \label{eq:master}
\end{equation}
where $\omega$ is the complex frequency when dealing with quasinormal modes and real frequency when solving the scattering problem, and $V_{s}(r)$ is the corresponding effective potential. Below we list the potentials for the (massive) scalar, electromagnetic (Maxwell), and massless Dirac fields.

\paragraph*{(i) Scalar field ($s=0$).}
A minimally coupled scalar of mass $\mu$ satisfies the Klein--Gordon equation,
\begin{equation}
  \Box\Phi-\mu^{2}\Phi=0,\qquad \Box\equiv g^{\mu\nu}\nabla_{\mu}\nabla_{\nu}.
\end{equation}
With the standard ansatz
\begin{equation}
  \Phi(t,r,\theta,\phi)=\frac{\Psi(r)}{r}\,Y_{\ell m}(\theta,\phi)\,e^{-i\omega t},
\end{equation}
one obtains Eq.~\eqref{eq:master} with
\begin{equation}
  V_{0}(r)=f(r)\!\left[\mu^{2}+\frac{\ell(\ell+1)}{r^{2}}+\frac{f'(r)}{r}\right],
  \qquad \ell=0,1,2,\ldots,
  \label{eq:V_scalar}
\end{equation}
where a prime denotes $d/dr$. The $\mu\to 0$ limit reproduces the usual massless scalar potential. Notice that for spacetimes which are asymptotically de Sitter a conformal coupling is usually introduced \cite{Tagirov:1972vv}, which, up to redefinition of constants, could be modelled by a massive term. 

\paragraph*{(ii) Electromagnetic field ($s=1$).}
Maxwell perturbations on a static, spherically symmetric background split into axial and polar sectors which are isospectral  on the spherically symmetric metric. After the usual vector-harmonic decomposition, both sectors reduce to Eq.~\eqref{eq:master} with
\begin{equation}
  V_{1}(r)=f(r)\,\frac{\ell(\ell+1)}{r^{2}},
  \qquad \ell=1,2,3,\ldots.
  \label{eq:V_EM}
\end{equation}
Thus the Maxwell potential is purely centrifugal and entirely governed by the background $f(r)$ and the angular barrier.

\paragraph*{(iii) Massless Dirac field ($s=1/2$).}
In the tetrad formalism, the curved-space Dirac equation
\begin{equation}
  \gamma^{\hat{\alpha}}\,e_{\hat{\alpha}}{}^{\mu}\!\left(\partial_{\mu}+\Gamma_{\mu}\right)\Upsilon=0
\end{equation}
can be separated using spinor spherical harmonics. Defining the superpotential
\begin{equation}
  W(r)=\Bigl(\ell+\tfrac{1}{2}\Bigr)\frac{\sqrt{f(r)}}{r},\qquad \ell=\tfrac{1}{2},\tfrac{3}{2},\tfrac{5}{2},\ldots,
\end{equation}
the two radial components satisfy supersymmetric partner equations of the form~\eqref{eq:master} with
\begin{equation}
  V_{1/2}^{(\pm)}(r)=W(r)^{2}\pm\frac{dW}{dr_{*}}
  =\Bigl(\ell+\tfrac{1}{2}\Bigr)^{2}\frac{f(r)}{r^{2}}
    \pm\Bigl(\ell+\tfrac{1}{2}\Bigr)\frac{1}{2r}\,\frac{df}{dr},
  \label{eq:V_Dirac}
\end{equation}
where the $\pm$ label the axial/polar (or “chiral”) sectors. The pair $V_{1/2}^{(+)}$ and $V_{1/2}^{(-)}$ is isospectral, so it suffices to analyze either one.

The potentials \eqref{eq:V_scalar}--\eqref{eq:V_Dirac} are short-ranged in the tortoise coordinate on the brane-projected Schwarzschild--de~Sitter background: in particular, $V_{s}\to 0$ as $r\to r_{h}$ (black-hole horizon, $r_{*}\to -\infty$) and also $V_{s}\to 0$ as $r\to r_{c}$ (cosmological horizon, $r_{*}\to +\infty$). These properties ensure that the scattering/QNM problems are well posed with the standard horizon boundary conditions used throughout the paper.

\section{Grey-body factors}\label{sec:GBF}

\begin{figure*}
\resizebox{\linewidth}{!}{
\includegraphics{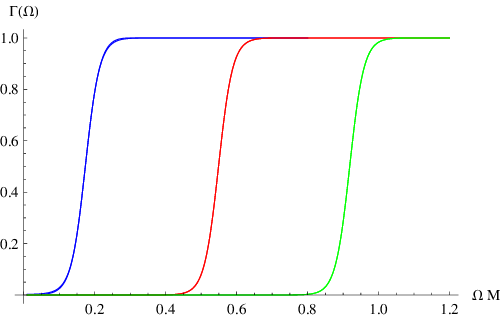}
\includegraphics{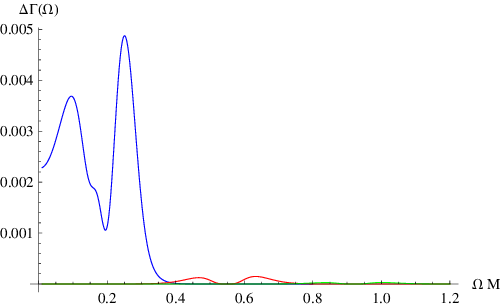}
}
\caption{Grey-body factors of the scalar field calculated by the 6th order WKB technique and by the correspondence with QNMs (left). The difference between the results obtained by the two methods (right). Here $M=1$, $s=0$, $\ell=0$, $\Lambda =0.7$, $\mu=1$ (blue), $\mu=3$ (red), $\mu=5$ (green); $D=5$.}\label{fig:GBL0s0}
\end{figure*}

\begin{figure*}
\resizebox{\linewidth}{!}{
\includegraphics{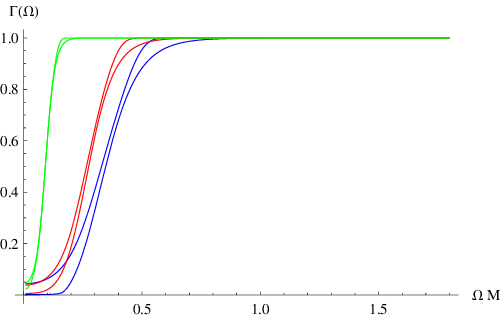}
\includegraphics{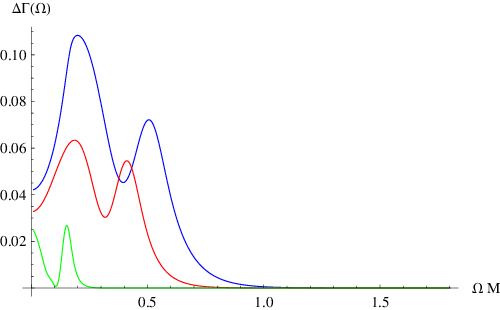}
}
\caption{Grey-body factors of the Dirac field calculated by the 6th order WKB technique and by the correspondence with QNMs (left). The difference between the results obtained by the two methods (right). Here $M=1$, $s=1/2$, $\ell=1/2$, $\Lambda=0$ (blue), $\Lambda=0.3$ (red), $\Lambda=0.7$ (green); $D=5$.}\label{fig:GBL12s12}
\end{figure*}

\begin{figure*}
\resizebox{\linewidth}{!}{
\includegraphics{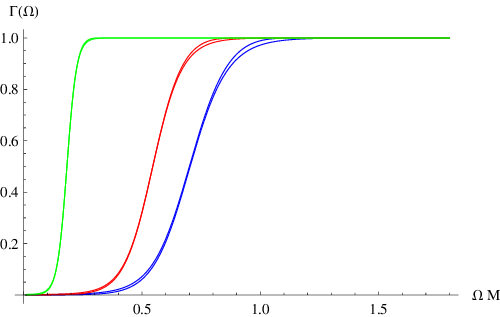}
\includegraphics{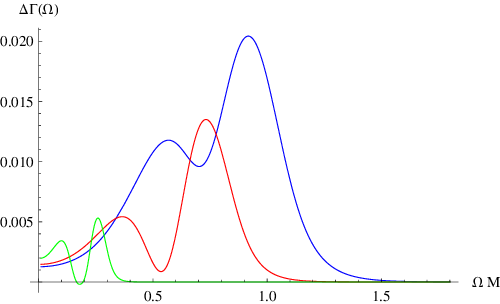}
}
\caption{Grey-body factors of the Dirac field calculated by the 6th order WKB technique and by the correspondence with QNMs (left). The difference between the results obtained by the two methods (right). Here $M=1$, $s=1/2$, $\ell=3/2$, $\Lambda=0$ (blue), $\Lambda=0.3$ (red), $\Lambda=0.7$ (green); $D=5$.}\label{fig:GBL32s12}
\end{figure*}

\begin{figure*}
\resizebox{\linewidth}{!}{
\includegraphics{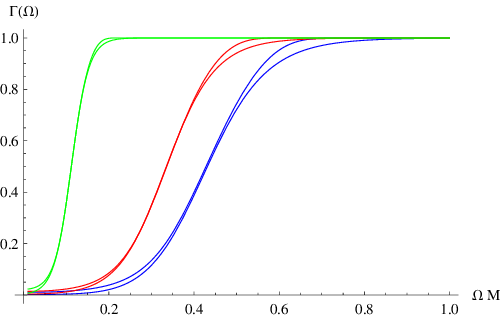}
\includegraphics{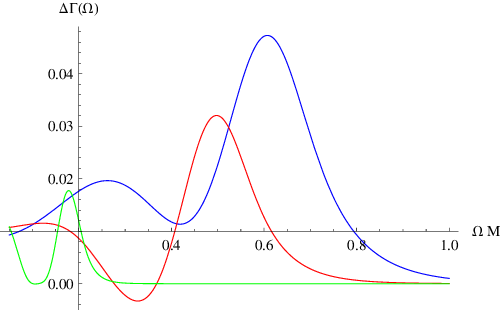}
}
\caption{Grey-body factors of the electromagnetic field calculated by the 6th order WKB technique and by the correspondence with QNMs (left). The difference between the results obtained by the two methods (right). Here $M=1$, $s=1$, $\ell=1$, $\Lambda=0$ (blue), $\Lambda=0.3$ (red), $\Lambda=0.7$ (green); $D=5$.}\label{fig:GBL1s1}
\end{figure*}

\begin{figure*}
\resizebox{\linewidth}{!}{
\includegraphics{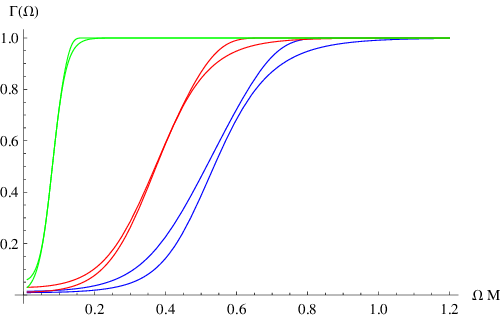}
\includegraphics{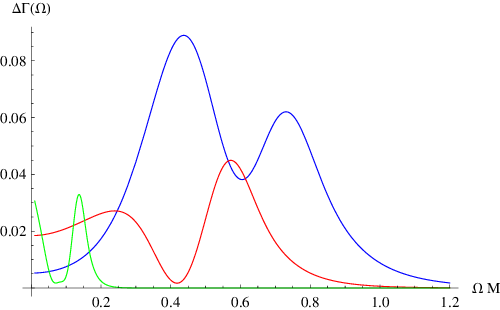}
}
\caption{Grey-body factors of the electromagnetic field calculated by the 6th order WKB technique and by the correspondence with QNMs (left). The difference between the results obtained by the two methods (right). Here $M=1$, $s=1$, $\ell=1$, $\Lambda=0$ (blue), $\Lambda=1$ (red), $\Lambda=2$ (green); $D=6$.}\label{fig:GBL1s1D6}
\end{figure*}

\begin{figure*}
\resizebox{\linewidth}{!}{
\includegraphics{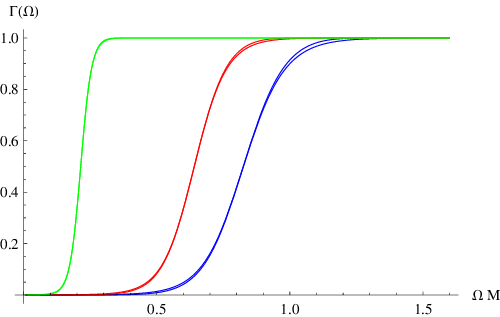}
\includegraphics{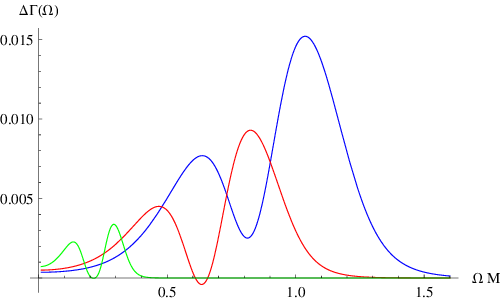}
}
\caption{Grey-body factors of the electromagentic field calculated by the 6th order WKB technique and by the correspondence with QNMs (left). The difference between the results obtained by the two methods (right). Here $M=1$, $s=1$, $\ell=2$, $\Lambda=0$ (blue), $\Lambda=0.3$ (red), $\Lambda=0.7$ (green); $D=5$.}\label{fig:GBL2s1}
\end{figure*}

\begin{figure*}
\resizebox{\linewidth}{!}{
\includegraphics{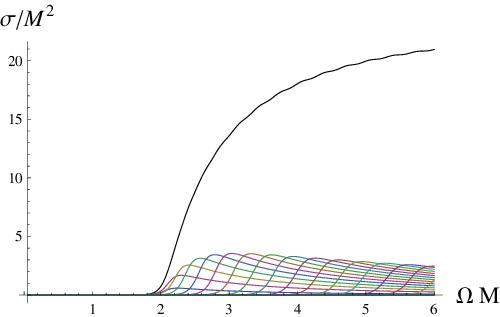}
\includegraphics{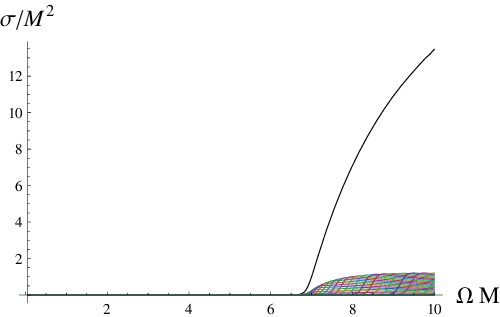}
}
\caption{Absorption cross-section for the first 25 multipoles of the scalar field together with the sum over partial cross-sections (top line). Here $M=1$, $s=0$, $\ell=0, 1, 2,...24$, $\Lambda=0.7$, $\mu=3$ (left), $\mu=10$ (right); $D=5$.}\label{fig:CSs0}
\end{figure*}

\begin{figure*}
\resizebox{\linewidth}{!}{
\includegraphics{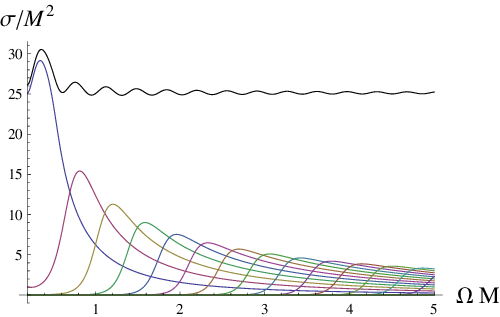}
\includegraphics{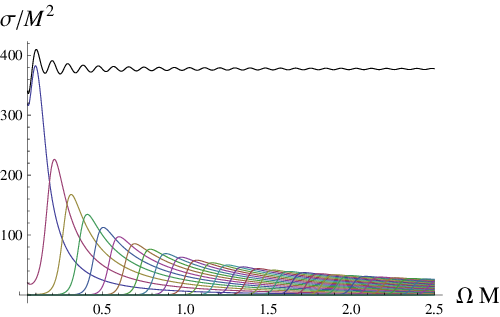}
}
\caption{Absorption cross-section for the first 25 multipoles of the Dirac field together with the sum over partial cross-sections (top line). Here $M=1$, $s=1/2$, $\ell=1/2, 3/2,...$, $\Lambda=0$ (left), $\Lambda=0.7$ (right); $D=5$.}\label{fig:CS12s12}
\end{figure*}

\begin{figure*}
\resizebox{\linewidth}{!}{
\includegraphics{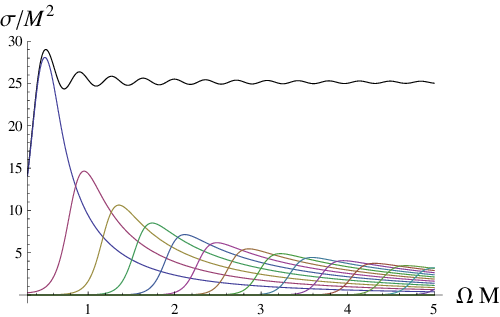}
\includegraphics{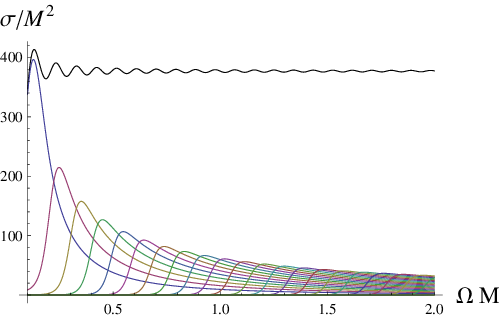}
}
\caption{Absorption cross-section for the first 25 multipoles of the electromagnetic  field together with the sum over partial cross-sections (top line). Here $M=1$, $s=1$, $\ell=1, 2, 3,...25$, $\Lambda=0$ (left), $\Lambda=0.7$ (right); $D=5$.}\label{fig:CSs1}
\end{figure*}

\begin{figure*}
\resizebox{\linewidth}{!}{
\includegraphics{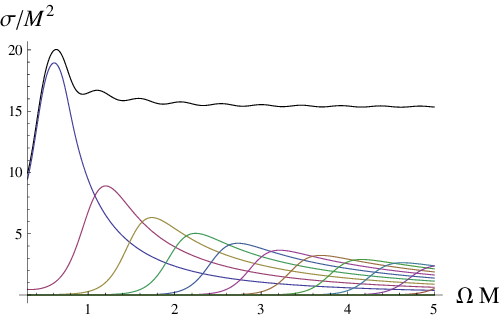}
\includegraphics{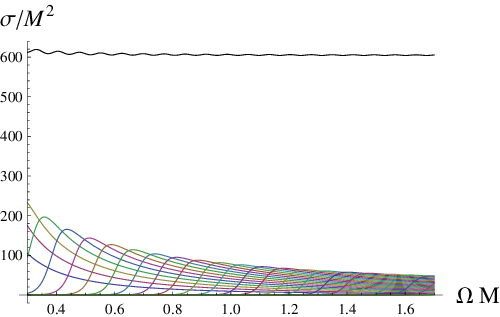}
}
\caption{Absorption cross-section for the first 25 multipoles of the electromagnetic  field together with the sum over partial cross-sections (top line). Here $M=1$, $s=1$, $\ell=1, 2, 3,...25$, $\Lambda=0$ (left), $\Lambda=2$ (right); $D=6$.}\label{fig:CSs1D6}
\end{figure*}

For each spin sector, the master equation for the perturbation can be written as
\begin{equation}
\frac{d^{2}\Psi}{dr_*^{2}}+\bigl(\Omega^{2}-V_\ell(r)\bigr)\Psi=0,
\qquad 
dr_*=\frac{dr}{f(r)},
\end{equation}
where $\Omega$ is the real frequency of the scattered wave and $V_\ell(r)$ is the effective potential for the given field.  
The potential $V_\ell(r)$ forms a single smooth barrier between the event horizon ($r_* \to -\infty$) and the cosmological horizon ($r_* \to +\infty$).

Scattering for real $\Omega>0$ is defined by the following boundary conditions:
\begin{widetext}
\begin{equation}
\Psi \sim 
\begin{cases}
T_\ell(\Omega)\,e^{-i\Omega r_*}, & r_* \to -\infty \quad (\text{purely ingoing at the black-hole horizon}),\\[2mm]
e^{-i\Omega r_*}+R_\ell(\Omega)\,e^{+i\Omega r_*}, & r_* \to +\infty \quad (\text{scattered wave in the far zone}),
\end{cases}
\end{equation}
so that the reflection and transmission coefficients satisfy current conservation,
\begin{equation}
|T_\ell(\Omega)|^{2}+|R_\ell(\Omega)|^{2}=1.
\end{equation}
\end{widetext}

The {\it grey-body factor} is defined as the transmission probability,
\begin{equation}
\Gamma_\ell(\Omega)=|T_\ell(\Omega)|^{2}=1-|R_\ell(\Omega)|^{2}.
\end{equation}
Let $r_{*0}$ denote the position of the maximum of $V_\ell(r)$, with
\begin{equation}
V_0\equiv V_\ell(r_{*0}), 
\qquad 
V_0^{(n)}\equiv \left.\frac{d^{n}V_\ell}{dr_*^{n}}\right|_{r_{*0}}.
\end{equation}
The $N$-th order Wentzel–Kramers–Brillouin (WKB) approximation \cite{Schutz:1985km,Iyer:1986np,Konoplya:2003ii,Matyjasek:2017psv,Konoplya:2019hlu}
gives the transmission probability in the compact form:
\begin{equation}
\boxed{
\Gamma_\ell^{\text{WKB}}(\Omega)
=\frac{1}{1+\exp\!\bigl(2\pi\,K_N(\Omega)\bigr)}
},
\end{equation}
where the WKB ``phase'' $K_N(\Omega)$ is
\begin{equation}
K_N(\Omega)=
\frac{\Omega^{2}-V_0}{\sqrt{-2V_0^{(2)}}}
-\sum_{k=2}^{N}\Lambda_k\!\left(\Omega;V_0^{(2)},V_0^{(3)},\ldots,V_0^{(2k)}\right),
\end{equation}
and the $\Lambda_k$ are known rational functions of the derivatives $V_0^{(j)}$ and of $(\Omega^{2}-V_0)$ (explicit forms up to high orders are given in \cite{Iyer:1986np,Matyjasek:2017psv}). The above relation was used for finding grey-body factors in numerous publications (see \cite{Dubinsky:2025nxv,Dubinsky:2024nzo,Konoplya:2021ube,Pathrikar:2025gzu} for recent applications).

In practical calculations:
\begin{enumerate}
\item One locates $r_{*0}$ (or equivalently $r_0$) such that $dV_\ell/dr|_{r_0}=0$;
\item Evaluates $V_0$ and its derivatives $V_0^{(j)}$ with respect to $r_*$;
\item Inserts them into $K_N(\Omega)$ and computes $\Gamma_\ell(\Omega)$ from the above formula.
\end{enumerate}

The Wentzel–Kramers–Brillouin (WKB) approach has established itself as one of the most versatile semi-analytic techniques for studying both quasinormal spectra and grey-body factors of black holes. Over the years, it has been successfully employed in a broad spectrum of scenarios — from higher-dimensional and asymptotically de Sitter spacetimes to black holes surrounded by exotic matter fields or arising in modified theories of gravity — providing reliable results even in geometries with nontrivial coupling between matter and curvature. 
~\cite{Dubinsky:2024mwd,Dubinsky:2025bvf,Dubinsky:2024rvf,Konoplya:2022hll,Malik:2024sxv,Dubinsky:2025fwv,Guo:2020blq,Malik:2024elk,Paul:2023eep,Konoplya:2001ji,Malik:2023bxc,Zinhailo:2019rwd,Fu:2022cul,Tan:2022vfe,Konoplya:2005sy,Churilova:2021tgn,DuttaRoy:2022ytr,Malik:2024tuf,Yu:2022yyv,Skvortsova:2023zmj,Skvortsova:2024eqi,Skvortsova:2024atk,Bolokhov:2025egl,Skvortsova:2025cah,Bolokhov:2023bwm,Bolokhov:2023ruj,Bolokhov:2024ixe}.

The WKB method is highly accurate for single-peaked, smooth potentials and for $\ell\gtrsim1$.  
It applies equally well to asymptotically flat and de Sitter spacetimes since, in both cases, $V_\ell(r)\to0$ as $r_*\to\pm\infty$, ensuring plane-wave asymptotics.
For massive fields, the mass modifies the shape of $V_\ell(r)$ but not the asymptotic form of the boundary conditions, so the same method remains valid.
However, accuracy decreases when the potential develops has no peak, which occurs for asymptotically flat case and larger mass of the field $\mu$. For asymptotically de Sitter case, the mass term produce the single peak potential and the WKB method usually provide good accuracy.

The computed grey-body factors (GBFs) for scalar, electromagnetic, and Dirac perturbations reveal the characteristic behavior expected from wave scattering by a potential barrier surrounding a black hole (see figs. 1-6). For each multipole number $\ell$, the transmission probability $\Gamma_{\ell}(\Omega)$ is strongly suppressed at low frequencies, increases rapidly through the transition region where the wave energy becomes comparable to the height of the effective potential, and asymptotically approaches unity in the high-frequency limit. This pattern reflects the tunneling nature of the process: low-frequency modes are reflected almost entirely, while high-frequency modes overcome the barrier with negligible reflection. Increasing $\ell$ shifts the onset of significant transmission to larger values of $\Omega$ and sharpens the transition, consistent with the fact that higher multipoles experience a stronger centrifugal barrier. When the partial contributions are summed over $\ell$, the individual oscillations are largely averaged out, producing a smooth total grey-body spectrum (see figs. 7-10). In this sum, the low-frequency absorption is dominated by the lowest available multipole (scalar: $\ell=0$, Dirac: $\ell=1/2$, electromagnetic: $\ell=1$), whereas higher $\ell$ contribute predominantly at larger frequencies.

The influence of the cosmological constant $\Lambda$ on the GBFs is clearly visible in the numerical results. Increasing $\Lambda$ reduces the height and width of the effective potential barrier between the black-hole and cosmological horizons. As a consequence, the transmission probabilities are enhanced for all field types, particularly at low and intermediate frequencies where tunneling dominates. Physically, the de Sitter expansion weakens the gravitational trapping near the photon sphere, making the potential more transparent to radiation. This trend is monotonic for both scalar and Dirac fields, while for electromagnetic perturbations the enhancement is somewhat weaker due to their larger effective barrier. The limit $\Lambda \to 0$ smoothly reproduces the familiar Schwarzschild behavior.

The field mass $\mu$ introduces an additional qualitative modification. As $\mu$ grows, the effective potential acquires a long-range positive tail, producing an extra “mass wall’’ at large radii that prevents low-frequency modes from escaping to infinity. Consequently, the grey-body factors for small $\Omega$ decrease rapidly with increasing $\mu$, and the transmission window shifts toward higher frequencies, roughly following the threshold $\Omega \simeq \mu$. In this regime, the total emission and absorption are strongly suppressed.

When the GBFs for all $\ell$ are combined, the total transmission curves interpolate between two limiting regimes: for $\Omega \ll 1$, the absorption is almost completely suppressed, reflecting the dominance of the potential barrier; for $\Omega \gg 1$, the transmission tends to unity, corresponding to the geometric-optics limit in which the wavelength is much shorter than the curvature scale. The intermediate region exhibits smooth peaks associated with the interplay between the effective potential and the wave frequency. These spectral features are sensitive to $\Lambda$, $\mu$, and $\ell$, providing a quantitative measure of how the cosmological expansion and field mass alter the scattering of radiation in brane-localized Schwarzschild–de Sitter black-hole spacetimes. Overall, the results demonstrate that both increasing $\Lambda$ and decreasing $\mu$ act to enhance the transparency of the potential barrier and thus the effective transmission of radiation from the black-hole vicinity.

\section{Correspondence between Grey-Body Factors and Quasinormal Modes}
\label{sec:correspondence}

The grey-body factors could be found once the least damped (fundamental) quasinormal mode $\omega_0$ and the first overtone $\omega_1$ are known. The quasinormal modes are proper oscillation frequencies under the boundary conditions requiring purely outgoing waves at infinity and purely incoming at the event horizon ~\cite{Kokkotas:1999bd,Berti:2009kk,Konoplya:2011qq,Bolokhov:2025uxz}. Then, the grey-body factors can be found via the following relation \cite{Konoplya:2024lir,Konoplya:2024vuj}
\[\begin{array}{r}
\Gamma_{\ell}(\Omega) =
\left[
1 + 
\exp\!\left(
\dfrac{2\pi\bigl[\Omega^{2} - \mathrm{Re}(\omega_{0})^{2}\bigr]}
     {4\,\mathrm{Re}(\omega_{0})\,\mathrm{Im}(\omega_{0})}
\right)
\right]^{-1}\\
+ \text{beyond eikonal corrections}.
\end{array}
\]
This correspondence has been tested and applied in a number of recent publications \cite{Han:2025cal,Dubinsky:2025nxv,Konoplya:2024lch,Malik:2024cgb,Lutfuoglu:2025hjy,Lutfuoglu:2025blw,Lutfuoglu:2025ldc,Bolokhov:2024otn,Malik:2025erb,Skvortsova:2024msa,Bolokhov:2025lnt}.
The correspondence evidently does not hold for the cases in which the WKB fails to reproduce the proper quasinormal modes \cite{Konoplya:2017wot,Bolokhov:2023dxq,Konoplya:2022gjp}. These include some theories with higher curvature corrections \cite{Konoplya:2020bxa,Konoplya:2025afm} where the eikonal instability takes place \cite{Dotti:2004sh,Gleiser:2005ra,Cuyubamba:2016cug,Konoplya:2017zwo,Konoplya:2017ymp}
.

The numerical analysis shown in figs. 1-6 confirms that the correspondence between grey-body factors and quasinormal modes holds with good accuracy for most of the parameter space, though its precision depends strongly on the multipole number $\ell$. 
For the monopole and dipole modes ($\ell=0,1$), the deviations between the grey-body factors obtained directly from the WKB transmission formula and those reconstructed via the correspondence are rather pronounced, reaching several percent in some cases. This reflects the limited applicability of the WKB approximation at low $\ell$, where the effective potential is broad and asymmetric, and the single-barrier assumption is only approximate. However, a crucial aspect of the correspondence’s validity lies in the fact that it applies only to the Schwarzschild branch of modes. The quasinormal spectrum of asymptotically de Sitter black holes consists of two distinct families: the Schwarzschild branch, which is perturbative in the cosmological constant $\Lambda$~\cite{Zhidenko:2003wq,Konoplya:2007zx,Konoplya:2013sba,Cuyubamba:2016cug,Dyatlov:2010hq,Jansen:2017oag,Molina:2003ff}, and the de Sitter branch, whose frequencies continuously approach those of pure de Sitter spacetime in the limit where the ratio of the black-hole radius to the de Sitter radius tends to zero~\cite{Lopez-Ortega:2012xvr,Lopez-Ortega:2007vlo}.

For $\ell=2$, the relative difference $\Delta\Gamma_{\ell}$ typically falls below one percent, and for $\ell \geq 3$ it becomes almost negligible. Thus, the correspondence can be regarded as quantitatively reliable already from the quadrupole mode onward, improving systematically with increasing $\ell$. The same tendency is evident from the plots of $\Delta\Gamma_{\ell}$, where the discrepancy decreases monotonically with higher multipole number.

The dependence on the background parameters is comparatively weak but still noticeable. Increasing the cosmological constant $\Lambda$ slightly enlarges the deviation, as the potential barrier becomes shallower and broader, while increasing the field mass $\mu$ tends to reduce the error by sharpening the barrier and improving the local WKB approximation. Overall, the results demonstrate that the correspondence between grey-body factors and quasinormal modes is accurate to within one percent for $\ell \geq 2$, while for the lowest multipoles it should be used with caution.

\section{Absorption Cross-Section}
\label{sec:absorption}

The absorption cross-section quantifies the effective area of the black hole that interacts with an incoming flux of radiation. Physically, it measures the probability that a wave impinging on the black hole will be absorbed rather than scattered back to infinity, and it plays a central role in characterizing black-hole emission and Hawking radiation spectra. In particular, it determines the rate at which different field modes contribute to the total energy flux reaching an observer at infinity.

For a given field of spin $s$ and frequency $\Omega$, the absorption cross-section can be expressed in terms of the grey-body factors $\Gamma_{\ell}(\Omega)$ as a multipole sum,
\begin{equation}
\sigma_{\text{abs}}(\Omega)
= \frac{\pi}{\Omega^{2}} \sum_{\ell=\ell_{\min}}^{\infty} (2\ell+1)\,\Gamma_{\ell}(\Omega),
\label{eq:sigmaabs}
\end{equation}
where $\ell_{\min}=0$ for scalar, $\ell_{\min}=1$ for electromagnetic, and $\ell_{\min}=1/2$ for Dirac fields.  
Each partial wave $(\ell,m)$ experiences the effective potential barrier described earlier, and the corresponding transmission probability $\Gamma_{\ell}(\Omega)$ determines the fraction of the incident flux that reaches the event horizon.  While the standard expression for the absorption cross section was originally derived for asymptotically flat spacetimes, it can be naturally extended to the de Sitter case by noting that the perturbation equation has the same Schrödinger-like form in terms of the tortoise coordinate $r_{*}$, which again spans from $-\infty$ to $+\infty$. The essential difference lies in the fact that, in the presence of a cosmological horizon, the physical region of wave propagation becomes finite, bounded by the event and cosmological horizons. When the cosmological constant is large, the de Sitter radius approaches the event horizon, and the available propagation region becomes increasingly compact.

At low frequencies, the absorption cross-section is dominated by the lowest multipoles and suppressed as $\Omega^{2}$, reflecting the increasing difficulty of low-energy waves to penetrate the barrier. In the opposite high-frequency (geometric-optics) regime, $\sigma_{\text{abs}}(\Omega)$ approaches the classical capture cross-section,  proportional to the area enclosed by the critical photon orbit. The intermediate-frequency region exhibits oscillatory behavior originating from interference between partial waves transmitted and reflected by the potential barrier. When summed over $\ell$, these oscillations partially average out, producing a smooth total absorption spectrum.

The behavior of $\sigma_{\text{abs}}$ reflects the black-hole parameters and the field mass $\mu$. Increasing $\Lambda$ typically reduces the effective potential height, consequently increases the grey-body factors, leading to an overall increase  of the absorption cross-section across all frequencies. Conversely, larger field mass $\mu$ introduces an additional long-range term in the potential that prevents low-frequency modes from escaping, thus shifting the onset of significant absorption to higher energies. The resulting suppression at small $\Omega$ becomes more pronounced with increasing $\mu$, especially for scalar fields (see figs. 7-10).

From an observational standpoint, the absorption cross-section determines the frequency-dependent transparency of the black-hole spacetime and is directly related to the emitted power in Hawking radiation. In this sense, it provides complementary information to the quasinormal spectrum, connecting the resonant and scattering aspects of black-hole perturbations within a unified framework.

\section{Conclusions}
\label{sec:conclusions}

In this work, we have investigated the propagation of test scalar, electromagnetic, and Dirac fields under scattering boundary conditions in the background of a higher-dimensional Schwarzschild–de Sitter black hole projected onto a 3+1-dimensional brane. Using the effective four-dimensional metric, we have derived the corresponding master wave equations and effective potentials, and analyzed the scattering and absorption properties of the black hole through the computation of grey-body factors and absorption cross-sections. The calculations were performed by means of the WKB method, which provides sufficiently accurate results in the regime where the effective potential has a single, well-defined barrier separating the event and cosmological horizons.

Our results show that the grey-body factors depend sensitively on the field spin and on the parameters of the background geometry. Increasing the cosmological constant $\Lambda$ lowers  the potential barrier, resulting in an overall  increasing of the grey-body factors across all frequencies.  For all fields, the low-frequency part of the spectrum is dominated by the lowest multipoles, while higher $\ell$ modes contribute only at larger frequencies where the potential barrier becomes transparent.  

The absorption cross-sections display the expected transition between  low- and high-frequency regimes: at small $\Omega$, absorption is strongly suppressed, while at large $\Omega$ it approaches the geometric-optics capture cross-section. The intermediate-frequency region is characterized by oscillations originating from interference between partial waves transmitted and reflected by the potential barrier. Increasing $\Lambda$ reduces both the height of the oscillations and the overall magnitude of the absorption cross-section, whereas the effect of increasing the field mass $\mu$ shifts the onset of significant absorption toward higher frequencies and suppresses the low-frequency tail.

We have also tested the correspondence between grey-body factors and quasinormal modes for the three types of perturbations. While the correspondence holds only approximately for the lowest multipoles ($\ell=0,1$), it becomes sufficiently accurate for $\ell\geq 2$, where the relative difference between the directly computed transmission coefficients and those reconstructed from quasinormal frequencies remains below one percent. This confirms the robustness of the eikonal correspondence in the regime where the WKB approximation is reliable.

Overall, the results presented here provide a consistent picture of how higher-dimensional and cosmological effects influence the transmission properties of black holes on the brane. They also demonstrate that the quasinormal–grey-body correspondence can be extended to brane-localized fields in asymptotically de Sitter geometries with reasonable precision. Possible extensions of this work include the analysis of charged and rotating brane-world black holes, as well as the study of polarization-dependent grey-body factors in more general higher-curvature or quantum-corrected scenarios.

\begin{acknowledgments}
The author thanks R. A. Konoplya for useful discussions. 
The author acknowledges the University of Seville for their support through the Plan-US of aid to Ukraine.
\end{acknowledgments}

\bibliography{bibliography}
\end{document}